\documentstyle[twocolumn,aps,epsf,prb]{revtex}

\begin{document}
\draft
\title{A one-map two-clock approach to teaching relativity in introductory physics}
\author{P. Fraundorf}
\address{Department of Physics \& Astronomy \\
University of Missouri-StL, \\
St. Louis MO 63121}
\date{\today }
\maketitle


\begin{abstract}
This paper presents some ideas which might assist teachers incorporating
special relativity into an introductory physics curriculum. One can define
the proper-time/velocity pair, as well as the coordinate-time/velocity pair,
of a traveler using only distances measured with respect to a single ``map''
frame. When this is done, the relativistic equations for momentum, energy,
constant acceleration, and force take on forms strikingly similar to their
Newtonian counterparts. Thus high-school and college students not ready for
Lorentz transforms may solve relativistic versions of any single-frame
Newtonian problems they have mastered. We further show that multi-frame
calculations (like the velocity-addition rule) acquire simplicity and/or
utility not found using coordinate velocity alone. From physics-9611011 
(xxx.lanl, NM, 1996).
\end{abstract}



\section{Introduction}

Since the 1920's, it has been known that classical Newtonian laws depart
from a description of reality, as velocity approaches the speed of light.
This divergence between Newtonian and relativistic physics was one of the
most remarkable discoveries of this century. Yet, more than 70 years later,
most American high schools and colleges teach introductory students
Newtonian dynamics, but then fail to teach students how to solve those same
problems relativistically. When relativity is discussed at all, it often
entails only a rather abstract introduction to Lorentz transforms, relegated
to the back of a chapter (or the back of a book).

This is in part due to the facts that: (i) Newton's laws work so well in
routine application, and (ii) introductions to special relativity (often
patterned after Einstein's introductions to the subject) focus on
discovery-philosophy rather than applications. Caution, and inertia
associated with old habits, may play a role as well. It is nonetheless
unfortunate because teaching Newtonian solutions without relativistic ones
at best leaves the student's education out-of-date, and deprives them of
experiences which might spark an interest in further physics education. At
worst, partial treatments may replace what is missing with misconceptions
about the complexity, irrelevance, and/or the limitations of relativity in
the study, for example, of simple things like uniform acceleration.

Efforts to link relativistic concepts to classical ones have been with us
from the beginning. For example, the observation that relativistic objects
behave at high speed as though their inertial mass increases in the $%
\overrightarrow{p}=m\overrightarrow{v}$ expression, led to the definition
(used in many early textbooks\cite{French}) of relativistic mass $m^{\prime
}\equiv m\gamma $. Such efforts are worthwhile because they can: (A)
potentially allow the introduction of relativity concepts at an earlier
stage in the education process by building upon already-mastered classical
relationships, and (B) find what is fundamentally true in both classical 
{\em and} relativistic approaches. The concepts of transverse ($m^{\prime }$%
) and longitudinal ($m^{\prime \prime }\equiv m\gamma ^3$) masses have
similarly been used\cite{Blatt} to preserve relations of the form $F_x=ma_x$
for forces perpendicular and parallel, respectively, to the velocity
direction.

Unfortunately for these relativistic masses, no deeper sense has emerged in
which the mass of a traveling object either changes, or has directional
dependence. Such masses allow familiar relationships to be used in keeping
track of non-classical behaviors (item A above), but do not (item B above)
provide frame-invariant insights or make other relationships simpler as
well. Hence majority acceptance of their use seems further away now\cite
{Adler} than it did several decades ago\cite{Goldstein}.

A more subtle trend in the literature has been toward the definition of a
quantity called proper velocity\cite{SearsBrehme,Shurcliff}, which can be
written as $\overrightarrow{w}\equiv \gamma \overrightarrow{v}$. We use the
symbol $w$ here because it is not in common use elsewhere in relativity
texts, and because $w$ resembles $\gamma v$ from a distance. This quantity
also allows the momentum expression above to be written in classical form as
a mass times a velocity, i.e. as $\overrightarrow{p}=m\overrightarrow{w}$.
Hence it serves one of the ``type A'' goals served by $m^{\prime }$ above.
However, it remains an interesting but ``homeless'' quantity in the present
literature. In other words, proper-velocity differs fundamentally from the
familiar coordinate-velocity, and unlike the latter has not in textbooks
been linked to a particular reference frame. After all, it uses distance
measured in an inertial frame but time measured on the clocks of a moving
and possibly accelerated observer. However, there is also something deeply
physical about proper-velocity. Unlike coordinate-velocity, it is a
synchrony-free\cite{Winnie,Ungar1} (i.e. local-clock only) means of
quantifying motion. Moreover, A. Ungar has recently made the case\cite
{Ungar2} that proper velocities, and not coordinate velocities, make up the
gyrogroup analog to the velocity group in classical physics. If so, then, is
it possible that introductory students might gain deeper physical insight
via its use?

The answer is yes. We show here that proper-velocity, when introduced as
part of a ``one-map two-clock'' set of time/velocity variables, allows us to
introduce relativistic momentum, time-dilation, {\em and} frame-invariant
relativistic acceleration/force into the classroom without invoking
discussion of multiple inertial frames or the abstract mathematics of
Lorentz transformation (item A above). Moreover, through use of
proper-velocity many relationships (including those like velocity-addition,
which require multiple frames or more than one ``map'') are made simpler and
sometimes more useful. When one simplification brings with it many others,
this suggests that ``item B'' insights may be involved as well. The three
sections to follow deal with the basic, acceleration-related, and multi-map
applications of this ``two-clock'' approach by first developing the
equations, and then discussing classroom applications.

\section{A traveler, one map, and two clocks}

One may argue that a fundamental break between classical and relativistic
kinematics involves the observation that time passes differently for moving
observers, than it does for stationary ones. In typical texts, discussion of
this fact involves discussion of separate traveler and map (e.g. primed and
unprimed) inertial reference frames, perhaps including Lorentz transforms
between them, even though the traveler may be accelerated and changing
reference frames constantly! This is not needed. Instead, we define two time
variables when describing the motion of a single object (or ``traveler'')
with respect to a single inertial coordinate frame (or ``map''). These time
variables are the ``map'' or coordinate-time $t$, and the ``traveler'' or
proper-time $\tau $. However, only one measurement of distances will be
considered, namely that associated with the inertial reference frame or
``map''.

It follows from above that two velocities will arise as well, namely the
coordinate-velocity $\overrightarrow{v}\equiv d\overrightarrow{x}/dt$, and
proper-velocity $\overrightarrow{w}\equiv d\overrightarrow{x}/d\tau $. The
first velocity measures map-distance traveled per unit map time, while the
latter measures map-distance traveled per unit traveler time. Each of these
velocities can be calculated from the other by knowing the
velocity-dependence of the ``traveler's speed of map-time'' $\gamma \equiv
dt/d\tau $, since it is easy to see from the definitions above that:

\begin{equation}
\overrightarrow{w}=\gamma \overrightarrow{v}\text{.}  \label{propervelocity}
\end{equation}

Because all displacements $dx$ are defined with respect to our map frame,
proper-velocity is not simply a coordinate-velocity measured with respect to
a different map. However, it does have a well-defined home, in fact with
many ``brothers and sisters'' who live there as well. This family is
comprised of the velocities reported by the infinite number of moving
observers who might choose to describe the motion of our traveler, with
their own clock on the map of their common ``home'' frame of reference\cite
{Noncoord}. One might call the members of this family ``non-coordinate
velocities'', to distinguish them from the coordinate-velocity measured by
an inertial observer who stays put in the frame of the map. The cardinal
rule for all such velocities is: {\em everyone measures displacements from
the vantage point of the home frame} (e.g. on a copy of a reference-frame
map in their own vehicle's glove compartment). Thus proper velocity $%
\overrightarrow{w}$ is that particular non-coordinate velocity which reports
the rate at which a given traveler's position on the reference map changes,
per unit time{\em \ on the clock of the traveler}.

\subsection{Developing the basic equations}

A number of useful relationships for the above ``traveler's speed of
map-time'' $\gamma $, including it's familiar relationship to
coordinate-velocity, follow simply from the nature of the flat spacetime
metric. Their derivation is outlined in Appendix A. For students not ready
for four-vectors, however, one can simply quote Einstein's prediction that
spacetime is tied together so that instead of $\gamma =1$, one has $\gamma
=1/\sqrt{1-(v/c)^2}=E/mc^2$, where $E$ is Einstein's ``relativistic energy''
and $c$ is the speed of light. By solving eqn. \ref{propervelocity} for $%
w(v) $, and putting the inverted solution $v(w)$ into the expression for $%
\gamma $ above, the following string of useful relationships follow
immediately:

\begin{eqnarray}
\gamma &\equiv &\frac{dt}{dt_o}  \nonumber \\
&=&\frac 1{\sqrt{1-\left( \frac vc\right) ^2}}=\sqrt{1+\left( \frac wc%
\right) ^2}=\frac E{mc^2}=1+\frac K{mc^2}\text{.}  \label{gamma}
\end{eqnarray}
Here of course $K$ is the kinetic energy of motion, equal classically to $%
\frac 12mv^2$.

Because equation \ref{gamma} allows one to relate velocities to energy, an
important part of relativistic dynamics is in hand as well. Another
important part of relativistic dynamics, mentioned in the introduction,
takes on familiar form since momentum at any speed is

\begin{equation}
\overrightarrow{p}=m\overrightarrow{w}=m\gamma \overrightarrow{v}\text{.}
\label{momentum}
\end{equation}
This relation has important scientific consequences as well. It shows that
momentum like proper velocity has no upper limit, and that coordinate
velocity becomes irrelevant to tracking momentum at high speeds (since for $%
w\gg c$, $v\simeq c$ and hence $p\propto $ $\gamma $). All of the equations
in this section are summarized for reference and comparison in Table \ref
{Table1}.

\subsection{Basic classroom applications}

One of the simplest exercises a student might perform is to show that, as
proper velocity $w$ goes to infinity, the coordinate-velocity $v$ never gets
larger than speed limit $c$. This can be done by simply solving equation \ref
{propervelocity} for $v$ as a function of $w$. Since student intuition
should argue strongly against ``map-distance per unit traveler time''
becoming infinite, an upper limit on coordinate-velocity $v$ may thus from
the beginning seem a very reasonable consequence. In typical introductory
courses, this upper limit on coordinate-velocity is not something students
are given a chance to prove for themselves. Students can also show, for
themselves at this point, that {\em classical} kinematics follows when all
speeds involved obey $v\ll c$, since this implies that $\gamma \simeq 1$
(cf. Table \ref{Table1}).

Given these tools to describe the motion of an object with respect to single
map frame, another type of relativistic problem within range is that of time
dilation. From the very definition of $\gamma $ as a ``traveler's speed of
map-time'', and the velocity relations which show that $\gamma \geq 1$, it
is easy for a student to see that the traveler's clock will always run
slower than map time. Hence if the traveler holds a fixed speed for a finite
time, one has from equation \ref{gamma} that traveler time is dilated
(spread out over a larger interval) relative to coordinate time, by the
relation

\begin{equation}
\Delta t=\gamma \Delta \tau \geq \Delta \tau  \label{dilation}
\end{equation}
Thus time-dilation problems can be addressed. This is one of several skills
that this strategy can offer to students taking only introductory physics,
an ``item A'' benefit according to the introduction. A {\em practical}
awareness of the non-global nature of time thus does not require readiness
for the abstraction of Lorentz transforms.

Convenient units for coordinate-velocity are [lightyears per map-year] or
[c]. Convenient units for proper-velocity, by comparison, are [lightyears
per traveler year] or [ly/tyr]. When proper-velocity reaches 1 [ly/tyr],
coordinate-velocity is $\frac 1{\sqrt{1+1}}=\frac 1{\sqrt{2}}\simeq 0.707$
[c]. Thus $w=1$ [ly/tyr] is a natural dividing line between classical and
relativistic regimes. In the absence of an abbreviation with mnemonic value
for 1 [ly/tyr], students sometimes call it a ``roddenberry'' [rb], perhaps
because in english this name evokes connections to ``hotrodding''
(high-speed), berries (minimal units for fruit), and a science fiction
series which ignores the lightspeed limit to which coordinate-velocity
adheres. It is also worth pointing out to students that, when measuring
times in years, and distances in light years, one earth gravity of
acceleration is conveniently $g\simeq 1.03$ [ly/yr$^2$].

We show here that the major difference between classical and two-clock
relativity involves the dependence of kinetic energy $K$ on velocity.
Instead of $\frac 12mv^2$, one has $mc^2(\sqrt{1+\left( \frac wc\right) ^2}%
-1)$ which by Taylor expansion in $\frac wc$ goes as $\frac 12mw^2$ when $%
w\ll c$. Although the relativistic expression is more complicated, it is not
prohibitive for introductory students, especially since they can first
calculate the physically interesting ``speed of map-time'' $\gamma $, and
then figure $K=mc^2(\gamma -1)$. If they are given rest-energy equivalents
for a number of common masses (e.g. for electrons $m_ec^2\simeq 511keV$),
this might make calculation of relativistic energies even less painful than
in the classical case!

Concerning momenta, one might imagine from its definition that
proper-velocity $w$ is the important speed to a relativistic traveler trying
to get somewhere on a map (say for example to Chicago) with minimum traveler
time. Equation \ref{momentum} shows that it is also a more interesting speed
from the point of view of law enforcement officials wishing to minimize
fatalities on futuristic highways where relativistic speeds are an option.
Proper velocity tells us what is physically important, since it is proportional
to the momentum available in the collision. If we want to ask how long it
will take an ambulance to get to the scene of an accident, then of course
coordinate velocity may be the key.

Given that proper velocity is the most direct link to physically important
quantities like traveler-time and momentum, it is not surprising that a
press unfamiliar with this quantity does not attend excitedly, for example,
to new settings of the ``land speed record'' for fastest accelerated
particle. New progress changes the value of $v$, the only velocity they are
prepared to talk about, in the 7th or 8th decimal place. The story of
increasing proper velocity, thus, goes untold to a public whose imagination
might be captured thereby. Thus proper-velocities for single $50GeV$
electrons in the LEP2 accelerator at CERN might be approaching $w=\gamma v=%
\frac E{mc^2}\times v\simeq \frac{50GeV}{511keV}\times c\simeq 10^5$
[lightyears per traveler year], while the educated lay public (comprised of
those who have had no more than an introductory physics course) is under a
vague impression that the lightspeed limit rules out major progress along
these lines.

\section{Acceleration and force with one map and two clocks.}

The foregoing relations introduce, in context of a single inertial frame and
without Lorentz transforms, many of the kinematical and dynamical relations
of special relativity taught in introductory courses, in modern physics
courses, and perhaps even in some relativity courses. In this section, we
cover less familiar territory, namely the equations of relativistic
acceleration. Forces if defined simply as rates of momentum-change in
special relativity have no frame-invariant formulation. That is, different
map frames will see different forces acting on a given traveler. Moreover,
solving problems with coordinate 3-vector acceleration alone can be very
messy, indeed\cite{French}.

Because of this, relativistic acceleration is seldom discussed in
introductory courses. Can relativistic equations for constant acceleration,
instead, be cast in familiar form? The answer is yes: a frame-invariant
3-vector acceleration, with simple integrals, arises naturally in one-map
two-clock relativity. Although the development of the (3+1)D equations is
tedious, we show that this acceleration bears a familiar relationship to the
frame-independent rate of momentum change (i.e. the force) {\em felt} by an
accelerated traveler.

\subsection{Developing the acceleration equations.}

By examining the frame-invariant scalar product of the acceleration
4-vector, one can show (as we do in the Appendix) that a ``proper
acceleration'' $\overrightarrow{\alpha }$ for our traveler, which is the
same to all inertial observers and thus ``frame-invariant'', can be written
in terms of components for the classical acceleration vector $%
\overrightarrow{a}$ by:

\begin{equation}
\overrightarrow{\alpha }=\frac{\gamma ^3}{\gamma _{\perp }}\overrightarrow{a}%
,\text{where }\overrightarrow{a}\equiv \frac{d^2\overrightarrow{x}}{dt^2}%
\text{.}  \label{properacceleration}
\end{equation}
This is remarkable, given that $\overrightarrow{a}$ is so strongly
frame-dependent! Here the ``transverse time-speed`` $\gamma _{\perp }$ is
defined as $1/\sqrt{1-(v_{\perp }/c)^2}$, where $v_{\perp }$ is the
component of coordinate velocity $\overrightarrow{v}$ perpendicular to the
direction of coordinate acceleration $\overrightarrow{a}$. In this section
generally, in fact, subscripts $\parallel $ and $\perp $ refer to parallel
and perpendicular component-directions relative to the direction of this
frame-invariant acceleration 3-vector $\overrightarrow{\alpha }$, and not
(for example) relative to coordinate velocity $\overrightarrow{v}$.

Before considering integrals of the motion for constant proper acceleration $%
\overrightarrow{\alpha }$, let's review the classical integrals of motion
for constant acceleration $\overrightarrow{a}$. These can be written as $%
a\simeq \Delta v_{\parallel }/\Delta t\simeq \frac 12\Delta (v^2)/\Delta
x_{\parallel }$. The first of these is associated with conservation of
momentum in the absence of acceleration, and the second with the work-energy
theorem. These may look more familiar in the form $v_{\parallel f}\simeq
v_{\parallel i}+a\Delta t$, and $v_{\parallel f}^2\simeq v_{\parallel
i}^2+2a\Delta x_{\parallel }$. Given that coordinate velocity has an upper
limit at the speed of light, it is easy to imagine why holding coordinate
acceleration constant in relativistic situations requires forces which
change even from the traveler's point of view, and is not possible at all
for $\Delta t>(c-v_{\parallel i})/a$.

Provided that proper time $\tau $, proper velocity $w$, and time-speed $%
\gamma $ can be used as variables, three simple integrals of the proper
acceleration can be obtained by a procedure which works for integrating
other non-coordinate velocity/time expressions as well\cite{Noncoord}. The
resulting integrals are summarized in compact form, like those above, as

\begin{equation}
\alpha =\gamma _{\perp }\frac{\Delta w_{\parallel }}{\Delta t}=c\frac{\Delta
\eta _{\parallel }}{\Delta \tau }=\frac{c^2}{\gamma _{\perp }}\frac{\Delta
\gamma }{\Delta x_{\parallel }}\text{.}  \label{integrals}
\end{equation}
Here the integral with respect to proper time $\tau $ has been simplified by
further defining the hyperbolic velocity angle or rapidity\cite
{TaylorWheeler} $\eta _{\parallel }\equiv \sinh ^{-1}[w_{\parallel
}/c]=\tanh ^{-1}[v_{\parallel }/c]$. Note that both $v_{\perp }$ and the
``transverse time-speed'' $\gamma _{\perp }$ are constants, and hence both
proper velocity, and longitudinal momentum $p_{\parallel }\equiv
mw_{\parallel }$, change at a uniform rate when proper acceleration is held
constant. If motion is only in the direction of acceleration, $\gamma
_{\perp }$ is 1, and $\Delta p/\Delta t=m\alpha $ in the classical tradition.

In classical kinematics, the rate at which traveler energy $E$ increases
with time is frame-dependent, but the rate at which momentum $p$ increases
is invariant. In special relativity, these rates (when figured with respect
to proper time) relate to each other as time and space components,
respectively, of the acceleration 4-vector. Both are frame-{\em dependent}
at high speed. However, we can define proper force separately as the force 
{\em felt} by an accelerated object. We show in the Appendix that this is
simply $\overrightarrow{F}\equiv m\overrightarrow{\alpha }$. That is, all
accelerated objects {\em feel} a frame-invariant 3-vector force $%
\overrightarrow{F}$ in the direction of their acceleration. The magnitude of
this force can be calculated from any inertial frame, by multiplying the
rate of momentum change {\em in the acceleration direction} times $\gamma
_{\perp }$, or by multiplying mass times the proper acceleration $\alpha $.
The classical relation $F\simeq dp/dt\simeq mdv/dt=md^2x/dt^2=ma$ then
becomes:

\begin{eqnarray}
F &=&\gamma _{\perp }\frac{dp_{\parallel }}{dt}=m\gamma _{\perp }\frac{%
dw_{\parallel }}{dt}=m\gamma _{\perp }\frac{d(\gamma v_{\parallel })}{dt} 
\nonumber \\
&=&m\frac{\gamma ^3}{\gamma _{\perp }}\frac{dv_{\parallel }}{dt}=m\frac{%
\gamma ^3}{\gamma _{\perp }}\frac{d^2x_{\parallel }}{dt^2}=m\frac{\gamma ^3}{%
\gamma _{\perp }}a=m\alpha  \label{properforce}
\end{eqnarray}
Even though the rate of momentum change joins the rate of energy change in
becoming frame-dependent at high speed, Newton's 2nd Law for 3-vectors thus
retains a frame-invariant form.

Although they depend on the observer's inertial frame, it is instructive to
write out the components of momentum and energy rate-of-change in terms of
proper force magnitude $F$. The classical equation relating rates of
momentum change to force is $d\overrightarrow{p}/dt\simeq \overrightarrow{F}%
\simeq ma\overrightarrow{i_{\parallel }}$, where $\overrightarrow{%
i_{\parallel }}$ is the unit vector in the direction of acceleration. This
becomes

\begin{equation}
\frac{d\overrightarrow{p}}{dt}=F\left[ \left( \frac 1{\gamma _{\perp }}%
\right) \overrightarrow{i_{\parallel }}+\left( \frac{\gamma _{\perp
}v_{\perp }}c\frac{v_{\parallel }}c\right) \overrightarrow{i_{\perp }}%
\right] \text{.}  \label{force}
\end{equation}
Note that if there are non-zero components of velocity in directions {\em %
both} parallel and perpendicular to the direction of acceleration, then
momentum changes are seen to have a component perpendicular to the
acceleration direction, as well as parallel to it. These transverse momentum
changes result because transverse proper velocity $w_{\perp }=\gamma
v_{\perp }$ (and hence momentum $p_{\perp }$) changes when traveler $\gamma $
changes, even though $v_{\perp }$ is staying constant.

As mentioned above, the rate at which traveler energy increases with time
classically depends on traveler velocity through the relation $dE/dt\simeq
Fv_{\parallel }\simeq m(\overrightarrow{a}\bullet \overrightarrow{v})$.
Relativistically, this becomes

\begin{equation}
\frac 1{\gamma _{\perp }}\frac{dE}{dt}=Fv_{\parallel }=m\left( 
\overrightarrow{\alpha }\bullet \overrightarrow{v}\right) \text{.}
\label{power}
\end{equation}
Hence the rate of traveler energy increase is in form very similar to that
in the classical case.

Similarly, the classical relationship between work, force, and impulse can
be summarized with the relation $dE/dx_{\parallel }\simeq \overrightarrow{F}%
\simeq dp_{\parallel }/dt$. Relativistically, this becomes

\begin{equation}
\frac 1{\gamma _{\perp }}\frac{dE}{dx_{\parallel }}=F=\gamma _{\perp }\frac{%
dp_{\parallel }}{dt}\text{.}  \label{WorkForceImpulse}
\end{equation}
Once again, save for some changes in scaling associated with the
``transverse time-speed'' constant $\gamma _{\perp }$, the form of the
classical relationship between work, force, and impulse is preserved in the
relativistic case. Since these simple connections are a result, and not the
reason, for our introduction of proper time/velocity in context of a single
inertial frame, we suspect that they provide insight into relations that are
true both classically and relativistically, and thus are benefits of ``type
B'' discussed in the introduction.

The development above is of course too complicated for an introductory
class. However, for the case of unidirectional motion, and constant
acceleration from rest, the Newtonian equations have exact relativistic
analogs except for the changed functional dependence of kinetic energy on
velocity. These equations are summarized in Table \ref{Table2}.

\subsection{Classroom applications of relativistic acceleration and force.}

In order to visualize the relationships defined by equation \ref{integrals},
it is helpful to plot for the (1+1)D or $\gamma _{\perp }=1$ case all
velocities and times versus $x$ in dimensionless form from a common origin
on a single graph (i.e. as $v/c$, $\alpha \tau /c$, $w/c=\alpha t/c$, and $%
\gamma $ versus $\alpha x/c^2$). As shown in Fig. 1, $v/c$ is asymptotic to
1, $\alpha \tau /c$ is exponential for large arguments, $w/c=\alpha t/c$ are
hyperbolic, and also tangent to a linear $\gamma $ for large arguments. The
equations underlying this plot, from \ref{integrals} for $\gamma _{\perp }=1$
and coordinates sharing a common origin, can be written simply as:

\begin{eqnarray}
\frac{\alpha x}{c^2}+1 &=&\sqrt{1+\left( \frac{\alpha t}c\right) ^2}=\cosh
\left[ \frac{\alpha \tau }c\right]  \nonumber \\
&=&\gamma =\frac 1{\sqrt{1-\left( \frac vc\right) ^2}}=\sqrt{1+\left( \frac w%
c\right) ^2}\text{.}  \label{2dplot}
\end{eqnarray}
This universal acceleration plot, adapted to the relevant range of
variables, can be used to illustrate the solution of, and possibly to
graphically solve, {\em any} constant acceleration problem. Similar plots
can be constructed for more complicated trips (e.g. accelerated twin-paradox
adventures) and for the (3+1)D case as well\cite{Noncoord}.

With plots of this sort, high school students can solve relativistic
acceleration problems with no equations at all! For example, consider
constant acceleration from rest at $\alpha =1$ [earth gravity] $\simeq 1$
[ly/yr$^2$] over a distance of 4 [lightyears]. One can read directly from
Fig.1 by drawing a line up from 4 on the x-axis that $\gamma \simeq 5$,
final proper-speed $w\simeq 4.8$ [ly/tyr], map-time $t\simeq 4.8$ [yr],
proper-time elapsed $\tau \simeq 2.3$ [yr], and final coordinate-speed $%
v\simeq 1$ [c]. Problems with most initial value sets can be solved
similarly, without equations, on such a plot with help from a straight edge
and a bit of trial and error. Of course, the range of variables involved
must be reflected in the ranges of the plot. For this reason, it may also
prove helpful to replot Fig.1 on a logarithmic scale. As you can see here,
in the classical limit when $w<<1$ [ly/tyr], all variables except $\gamma $
(dimensionless times and velocities alike) {\em take on the same value} as a
function of distance traveled from rest!

For a numerical example, imagine trying to predict how far one might travel
by accelerating at one earth gravity for a fixed traveler-time, and then
turning your thrusters around and decelerating for the same traveler-time
until you are once more at rest in your starting or ``map'' frame. To be
specific, consider the 14.2 proper-year first half of such a trip all the
way to the Andromeda galaxy\cite{LagouteDavoust}, one of the most distant
(and largest) objects visible to the naked eye. From equation \ref{integrals}%
, the maximum (final) rapidity is simply $\eta _{\parallel }=\alpha \tau
/c=14.7$. Hence the final proper velocity is $w=\sinh (\alpha \tau
/c)=1.2\times 10^6ly/tyr$. From equation \ref{gamma} this means that $\gamma
=\sqrt{1+(w/c)^2}=1.2\times 10^6$, and the coordinate velocity $v=w/\sqrt{%
1+(w/c)^2}=0.99999999999963ly/yr$. Going back to equation \ref{integrals},
this means that coordinate time elapsed is $t=w/\alpha c=1.1\times 10^6years$%
, and distance traveled $x=(\gamma -1)c^2/\alpha =1.1\times 10^6ly$. Few
might imagine, from typical intro-physics treatments of relativity, that one
could travel over a million lightyears in less than 15 years on the
traveler's clock!

From equation \ref{properacceleration}, the coordinate acceleration falls
from $1gee$ at the start of the leg to $a=\alpha /\gamma ^3=6\times
10^{-19}gee$ at maximum speed. The forces, energies, and momenta of course
depend on the spacecraft's mass. At any given point along the trajectory
from the equations above, $F$ is of course just $m\alpha $, $dE/dx$ is $%
\gamma _{\perp }F=F$, $dp/dt$ is $F/\gamma _{\perp }=F$, and $dE/dt$ is $%
\gamma _{\perp }Fv_{\parallel }=Fv$. Note that all except the last of these
are constant if mass is constant, albeit dependent on the reference frame
chosen. However, the 4-vector components $dp/d\tau $ and $dE/d\tau $ are not
constant at all, showing in another way the pervasive frame-dependences
mentioned above.

The foregoing solution may seem routine, as well it should be. It is not.
Note that it was implemented using distances measured (and concepts defined)
in context of a single map frame. Moreover, the 3-vector forces and
accelerations used and calculated have frame-invariant components, i.e.
those particular parameters are correct in context of all inertial frames.

The mass of the ship in the problem above may vary with time. For example,
if the spacecraft is propelled by ejecting particles at velocity $u$
opposite to the acceleration direction, the force felt in the frame of the
traveler will be simply $m\alpha =-udm/d\tau $. Hence in terms of traveler
time the mass obeys $m=m_o\exp [-\alpha \tau /u]$. In terms of coordinate
time, the differential equation becomes $m\alpha =-u\gamma dm/dt$. This can
be solved to get the solution derived with significantly more trouble in the
reference above\cite{LagouteDavoust}.

\section{Problems involving more than one map.}

The foregoing sections treat calculations made possible, and analogies with
classical forms which result, if one introduces the proper time/velocity
variables in the context of a single map frame. What happens when multiple
map-frames are required? In particular, are the Lorentz transform and other
multi-map relations similarly simplified or extended? The answer is yes,
although our insights in this area are limited since the focus of this paper
is introductory physics, and not special relativity.

\subsection{Development of multi-map equations}

The Lorentz transform itself is simplified with the help of proper velocity,
in that it can be written in the symmetric matrix form:

\begin{equation}
\left[ 
\begin{array}{c}
c\Delta t^{\prime } \\ 
\Delta x^{\prime } \\ 
\Delta y^{\prime } \\ 
\Delta z^{\prime }
\end{array}
\right] =\left( 
\begin{array}{cccc}
\gamma & \pm \frac wc & 0 & 0 \\ 
\pm \frac wc & \gamma & 0 & 0 \\ 
0 & 0 & 1 & 0 \\ 
0 & 0 & 0 & 1
\end{array}
\right) \left[ 
\begin{array}{c}
c\Delta t \\ 
\Delta x \\ 
\Delta y \\ 
\Delta z
\end{array}
\right] \text{.}  \label{transform}
\end{equation}
This seems to be an improvement over the asymmetric equations normally used,
but of course requires a bit of matrix and 4-vector notation that your
students may not be ready to exploit.

The expression for length contraction, namely $L=L_o/\gamma $, is not
changed at all. The developments above do suggest that the concept of proper
length $L_o$, as the length of a yardstick in the frame in which it is at
rest, may have broader use as well. The relativistic Doppler effect
expression, given as $f=f_o\sqrt{\{1+(v/c)\}/\{1-(v/c)\}}$ in terms of
coordinate velocity, also simplifies to $f=f_o/\{\gamma -(w/c)\}$. The
classical expression for the Doppler effected frequency of a wave of
velocity $v_{wave}$ from a moving source of frequency $f_o$ is, for
comparison, $f=f_o/\{1-(v/v_{wave})\}$.

The most noticeable effect of proper velocity, on the multi-map
relationships considered here, involves simplification and symmetrization of
the velocity addition rule. The rule for adding coordinate velocities $%
\overrightarrow{v}^{\prime }$ and $\overrightarrow{v}$ to get relative
coordinate velocity $\overrightarrow{v}^{\prime \prime }$, namely $%
v_{\parallel }^{\prime \prime }=(v^{\prime }+v_{\parallel })/(1+v_{\parallel
}v^{\prime }/c^2$) and $v_{\perp }^{\prime \prime }\neq v_{\perp }$ with
subscripts referring to component orientation with respect to the direction
of $\overrightarrow{v}^{\prime }$, is inherently complicated. Moreover, for
high speed calculations, the answer is usually uninteresting since large
coordinate velocities always add up to something very near to $c$. By
comparison, if one adds proper velocities $\overrightarrow{w}^{\prime
}=\gamma ^{\prime }\overrightarrow{v}^{\prime }$ and $\overrightarrow{w}%
=\gamma \overrightarrow{v}$ to get relative proper velocity $\overrightarrow{%
w}^{\prime \prime }$, one finds simply that the coordinate velocity factors
add while the $\gamma $-factors multiply, i.e.

\begin{equation}
w_{\parallel }^{\prime \prime }=\gamma ^{\prime }\gamma \left( v^{\prime
}+v_{\parallel }\right) \text{,with }w_{\perp }^{\prime }=w_{\perp }\text{.}
\label{addition}
\end{equation}
Note that the components transverse to the direction of $\overrightarrow{v}%
^{\prime }$ are unchanged. These equations are summarized for the
unidirectional motion case in Table \ref{Table3}.

\subsection{Classroom applications involving more than one map-frame.}

Physically more interesting questions can be answered with equation \ref
{addition} than with the coordinate velocity addition rule commonly given to
students. For example, one might ask what the speed record is for {\em %
relative} proper velocity between two objects accelerated by man. For the
world record in this particle-based demolition derby, consider colliding two
beams from an accelerator able to produce particles of known energy for
impact onto a stationary target. From Table \ref{Table1} for colliding $%
50GeV $ electrons in the LEP2 accelerator at CERN, $\gamma $ and $\gamma
^{\prime } $ are $E/mc^2\simeq 50GeV/511keV\simeq 10^5$, $v$ and $v^{\prime }
$ are essentially $c$, and $w$ and $w^{\prime }$ are hence $10^5c$. Upon
collision, equation \ref{addition} tells us that the relative proper speed $%
w^{\prime \prime }$ is $(10^5)^2(c+c)=2\times 10^{10}c$. Investment in a
collider thus buys a factor of $2\gamma =2\times 10^5$ increase in the
momentum (and energy) of collision. Compared to the cost of building a $%
10PeV $ accelerator for the equivalent effect on a stationary target, the
collider is a bargain indeed!

\section{Conclusions.}

We show in this paper that a one-map two-clock approach, using both proper
and coordinate velocities, lets students tackle time dilation as well as
momentum and energy conservation problems without having to first master
concepts which arise when considering more than one inertial frame (like
Lorentz transforms, length contraction, and frame-dependent simultaneity).
The cardinal rule to follow when doing this is simple: All distances must be
defined with respect to a ``map'' drawn from the vantage point of a single
inertial reference frame.

We show further that a {\em frame-invariant} proper acceleration 3-vector
has three simple integrals of the motion in terms of these variables. Hence
students can speak of the proper acceleration and force 3-vectors for an
object in map-independent terms, and solve relativistic constant
acceleration problems much as they now do for non-relativistic problems in
introductory courses.

We have provided some examples of the use of these equations for high school
and college introductory physics classes, as well as summaries of equations
for the simple unidirectional motion case (Tables \ref{Table1}, \ref{Table2}%
, and \ref{Table3}). In the process, one can see that the approach does more
than ``superficially preserve classical forms''. Not just one, but many,
classical expressions take on relativistic form with only minor change. In
addition, interesting physics is accessible to students more quickly with
the equations that result. The relativistic addition rule for proper
velocities is a special case of the latter in point. Hence we argue that the
trend in the pedagogical literature, away from relativistic masses and
toward use of proper time and velocity in combination, may be a robust one
which provides: (B) deeper insight, as well as (A) more value from lessons
first-taught.

\acknowledgments

My thanks for input relevant to this paper from G. Keefe, W. A. Shurcliff,
E. F. Taylor, and A. A. Ungar. The work has benefited indirectly from
support by the U.S. Department of Energy, the Missouri Research Board, as
well as Monsanto and MEMC Electronic Materials Companies. It has benefited
most, however, from the interest and support of students at UM-St. Louis.

\appendix

\section{The 4-vector perspective}

This appendix provides a more elegant view of matters discussed in the body
of this paper by using space-time 4-vectors not used there, along with some
promised derivations. We postulate first that: {\bf (i)} displacements
between events in space and time may be described by a displacement 4-vector 
${\bf X}$ for which the time--component may be put into distance-units by
multiplying by the speed of light $c$; {\bf (ii)} subtracting the sum of
squares of space-related components of any 4-vector from the time component
squared yields a scalar ``dot-product'' which is {\em frame-invariant}, i.e.
which has a value which is the same for all inertial observers; and {\bf %
(iii)} translational momentum and energy, two physical quantities which are
conserved in the absence of external intervention, are components of the
momentum-energy 4-vector ${\bf P\equiv }m\frac{d{\bf X}}{d\tau }$, where m
is the object's rest mass and $\tau $ is the frame-invariant displacement in
time-units along its trajectory.

From above, the {\em 4-vector displacement} between two events in space-time
is described in terms of the position and time coordinate values for those
two events, and can be written as:

\begin{equation}
\Delta {\bf X}\equiv \left[ 
\begin{array}{c}
c\Delta t \\ 
\Delta x \\ 
\Delta y \\ 
\Delta z
\end{array}
\right] \text{.}  \label{displacement}
\end{equation}
Here the usual $\Delta $-notation is used to represent the value of final
minus initial. The dot-product of the displacement 4-vector is defined as
the square of the frame-invariant proper-time interval between those two
events. In other words,

\begin{eqnarray}
(c\Delta \tau )^2 &\equiv &\Delta {\bf X}\bullet \Delta {\bf X}  \nonumber \\
&=&(c\Delta t)^2-(\Delta x^2+\Delta y^2+\Delta z^2)\text{.}
\label{propertime}
\end{eqnarray}
Since this dot-product can be positive or negative, proper time intervals
can be real (time-like) or imaginary (space-like). It is easy to rearrange
this equation for the case when the displacement is infinitesimal, to
confirm the first two equalities in equation \ref{gamma} via:

\begin{equation}
\gamma \equiv \frac{dt}{d\tau }=\sqrt{1+\left( \frac{dx}{d\tau }\right) ^2}=%
\frac 1{\sqrt{1-\left( \frac{dx}{dt}\right) ^2}}\text{.}  \label{speedoftime}
\end{equation}

The {\em momentum-energy 4-vector}, as mentioned above, is then written
using $\gamma $ and the components of proper velocity $\overrightarrow{w}%
\equiv \frac{d\overrightarrow{x}}{d\tau }$ as:

\begin{equation}
{\bf P}\equiv m{\bf U}=m\left[ 
\begin{array}{c}
c\gamma \\ 
w_x \\ 
w_y \\ 
w_z
\end{array}
\right] =\left[ 
\begin{array}{c}
\frac Ec \\ 
p_x \\ 
p_y \\ 
p_z
\end{array}
\right] \text{.}  \label{momentumenergy}
\end{equation}
Here we've also taken the liberty to use a {\em velocity 4-vector} ${\bf U}%
\equiv d{\bf X}/d\tau $. The equality in equation \ref{gamma} between $%
\gamma $ and $E/mc^2$ follows immediately. The frame-invariant dot-product
of this 4-vector, times $c$ squared, yields the familiar relativistic
relation between total energy $E$, momentum $p$, and frame-invariant rest
mass-energy $mc^2$:

\begin{equation}
c^2{\bf P}\bullet {\bf P}=\left( mc^2\right) ^2=E^2-\left( cp\right) ^2\text{%
.}  \label{restenergy}
\end{equation}
If we define kinetic energy as the difference between rest mass-energy and
total energy using $K\equiv E-mc^2$, then the last equality in equation \ref
{gamma} follows as well. Another useful relation which follows is the
relation between infinitesimal uncertainties, namely $\frac{dE}{dp}=\frac{dx%
}{dt}$.

Lastly, the {\em force-power 4-vector} may be defined as the proper time
derivative of the momentum-energy 4-vector, i.e.:

\begin{equation}
{\bf F}\equiv \frac{d{\bf P}}{d\tau }\equiv m{\bf A}=m\left[ 
\begin{array}{c}
c\frac{d\gamma }{d\tau } \\ 
\frac{dw_x}{d\tau } \\ 
\frac{dw_y}{d\tau } \\ 
\frac{dw_z}{d\tau }
\end{array}
\right] =\left[ 
\begin{array}{c}
\frac 1c\frac{dE}{d\tau } \\ 
\frac{dp_x}{d\tau } \\ 
\frac{dp_y}{d\tau } \\ 
\frac{dp_z}{d\tau }
\end{array}
\right] \text{.}  \label{forcepower}
\end{equation}
Here we've taken the liberty to define {\em acceleration 4-vector} ${\bf A}%
\equiv d^2{\bf X}/d\tau ^2$ as well.

The dot-product of the force-power 4-vector is always negative. It may
therefore be used to define the frame-invariant proper acceleration $\alpha $%
, by writing: 
\begin{equation}
{\bf F}\bullet {\bf F}\equiv -\left( m\alpha \right) ^2=\left( \frac 1c\frac{%
dE}{d\tau }\right) ^2-\left( \frac{dp}{d\tau }\right) ^2\text{.}
\label{alpha}
\end{equation}
We still must show that this frame-invariant proper acceleration has the
magnitude specified in the text (eqn. \ref{properacceleration}). To relate
proper acceleration $\alpha $ to coordinate acceleration $\overrightarrow{a}%
\equiv \frac{d\overrightarrow{v}}{dt}\equiv \frac{d^2\overrightarrow{x}}{dt^2%
}$, note first that $c\frac{d\gamma }{d\tau }=\gamma ^4\frac{v_{\parallel }}c%
a$, that $\frac{dw_{\parallel }}{d\tau }=\frac{\gamma ^4}{\gamma _{\perp }^2}%
a$, and that $\frac{dw_{\perp }}{d\tau }=\gamma ^3\frac{v_{\perp }}c\frac{%
v_{\parallel }}ca$. Putting these results into the dot-product expression
for the fourth term in \ref{forcepower} and simplifying yields $\alpha ^2=%
\frac{\gamma ^6}{\gamma _{\perp }^2}a^2$ as required.

As mentioned in the text, power is classically frame-dependent, but
frame-dependence for the components of momentum change only asserts itself
at high speed. This is best illustrated by writing out the force 4-vector
components for a trajectory with constant proper acceleration, in terms of 
{\em frame-invariant} proper time/acceleration variables $\tau $ and $\alpha 
$. If we consider separately the momentum-change components parallel and
perpendicular to the unchanging and frame-independent acceleration 3-vector $%
\overrightarrow{\alpha }$, one gets

\begin{equation}
{\bf F}=\left[ 
\begin{array}{c}
\frac 1c\frac{dE}{d\tau } \\ 
\frac{dp_{\parallel }}{d\tau } \\ 
\frac{dp_{\perp }}{d\tau } \\ 
0
\end{array}
\right] =m\alpha \left[ 
\begin{array}{c}
\gamma _{\perp }\sinh \left[ \frac{\alpha \tau }c+\eta _o\right] \\ 
\cosh \left[ \frac{\alpha \tau }c+\eta _o\right] \\ 
\gamma _{\perp }\frac{v_{\perp }}c\sinh \left[ \frac{\alpha \tau }c+\eta
_o\right] \\ 
0
\end{array}
\right] \text{,}  \label{constant4acceleration}
\end{equation}
where $\eta _o$ is simply the initial value for $\eta _{\parallel }\equiv
\sinh ^{-1}[\frac{w_{\parallel }}c]$.

The force responsible for motion, as distinct from the frame-dependent rates
of momentum change described above, is that seen by the accelerated object
itself. As equation \ref{constant4acceleration} shows for $\tau ,v_{\perp }$
and $\eta _o$ set to zero, this is nothing more than $\overrightarrow{F}%
\equiv m\overrightarrow{\alpha }$. Thus some utility for the rapidity/proper
time integral of the equations of constant proper acceleration (3rd term in
eqn. \ref{integrals}) is illustrated as well.

\onecolumn

\begin{table}[tbp] \centering
\begin{tabular}{rcc}
{\bf Equation%
\mbox{$\backslash$}
Version:} & classical (c$\rightarrow \infty $) & two-clock relativity \\ 
\hline\hline
speed of map-time & $\gamma \equiv \frac{dt}{d\tau }\simeq 1$ & $\gamma
\equiv \frac{dt}{d\tau }=\frac 1{\sqrt{1-\left( \frac vc\right) ^2}}=\sqrt{%
1+\left( 
{\textstyle {w \over c}}
\right) ^2}$ \\ 
time dilation & none & $\Delta t=\gamma \Delta \tau $ \\ 
coordinate velocity & $\overrightarrow{v}\equiv \frac{d\overrightarrow{x}}{dt%
}$ & $\overrightarrow{v}\equiv \frac{d\overrightarrow{x}}{dt}=\frac{%
\overrightarrow{w}}{\sqrt{1+\left( \frac wc\right) ^2}}$ \\ 
proper velocity & same as coordinate & $\overrightarrow{w}\equiv \frac{d%
\overrightarrow{x}}{d\tau }=\gamma \overrightarrow{v}$ \\ \hline
momentum & $\overrightarrow{p}\simeq m\overrightarrow{v}$ & $\overrightarrow{%
p}=m\overrightarrow{w}$ \\ 
kinetic energy & $K\simeq \frac 12mv^2\simeq \frac{p^2}{2m}$ & $K=mc^2\left(
\gamma -1\right) \equiv E-mc^2$ \\ 
total energy $mc^2+K$ & not considered & $E=\gamma mc^2=\sqrt{\left(
pc\right) ^2+\left( mc^2\right) ^2}$ \\ \hline
\end{tabular}
\caption{Equations involving velocities, times, momentum and energy, 
in classical and two-clock relativistic form.
\label{Table1}}%
\end{table}

\begin{table}[tbp] \centering
\begin{tabular}{rcc}
{\bf Equation%
\mbox{$\backslash$}
Version:} & classical (c$\rightarrow \infty $) & two-clock relativity \\ 
\hline
coordinate acceleration & $a\equiv \frac{dv}{dt}$ & $a\equiv \frac{dv}{dt}$
\\ 
``felt'' or proper acceleration & same as coordinate & $\alpha =\frac{dw}{dt}%
=\gamma ^3a$ \\ \hline
momentum integral & $\frac pm\simeq at\simeq v$ & $\frac pm=\alpha t=w$ \\ 
work-energy integral & $\frac Km\simeq ax\simeq \frac 12v^2$ & $\frac Km%
=\alpha x=c^2(\gamma -1)$ \\ 
proper-time integral & same as momentum & $\alpha \tau =c\sinh ^{-1}\left( 
\frac wc\right) $ \\ \hline
force, work \& impulse & $F\simeq ma\simeq \frac{dE}{dx}=\frac{dp}{dt}$ & $%
F=m\alpha =\frac{dE}{dx}=\frac{dp}{dt}$ \\ \hline
\end{tabular}
\caption{Unidirectional motion equations involving constant acceleration from rest, 
and ``2nd law'' dynamics, in classical and two-clock relativistic form.\label{Table2}}%
\end{table}

\begin{table}[tbp] \centering
\begin{tabular}{rcc}
{\bf Equation%
\mbox{$\backslash$}
Version:} & classical (c$\rightarrow \infty $) & two-clock relativity \\ 
\hline
frame transformation & $x^{\prime }\simeq x\pm vt$; $t^{\prime }\simeq t$ & $%
x^{\prime }=\gamma x\pm wt$; $t^{\prime }=\gamma t\pm wx/c^2$ \\ 
length contraction & none & $L=L_o/\gamma $ \\ 
moving-source Doppler-shift & $f\simeq \frac{f_{source}}{1\pm v/v_{wave}}$
(any wave) & $f=\frac{f_{source}}{\gamma \pm w/c}$ (light) \\ 
velocity addition & $v_{ac}\simeq v_{ab}+v_{bc}$ & $w_{ac}\simeq \gamma
_{ab}\gamma _{bc}(v_{ab}+v_{bc})$ \\ \hline
\end{tabular}
\caption{Unidirectional motion equations involving distances measured
using more than one map-frame, in classical and two-clock relativistic form.
\label{Table3}}%
\end{table}

\begin{figure}
\epsffile{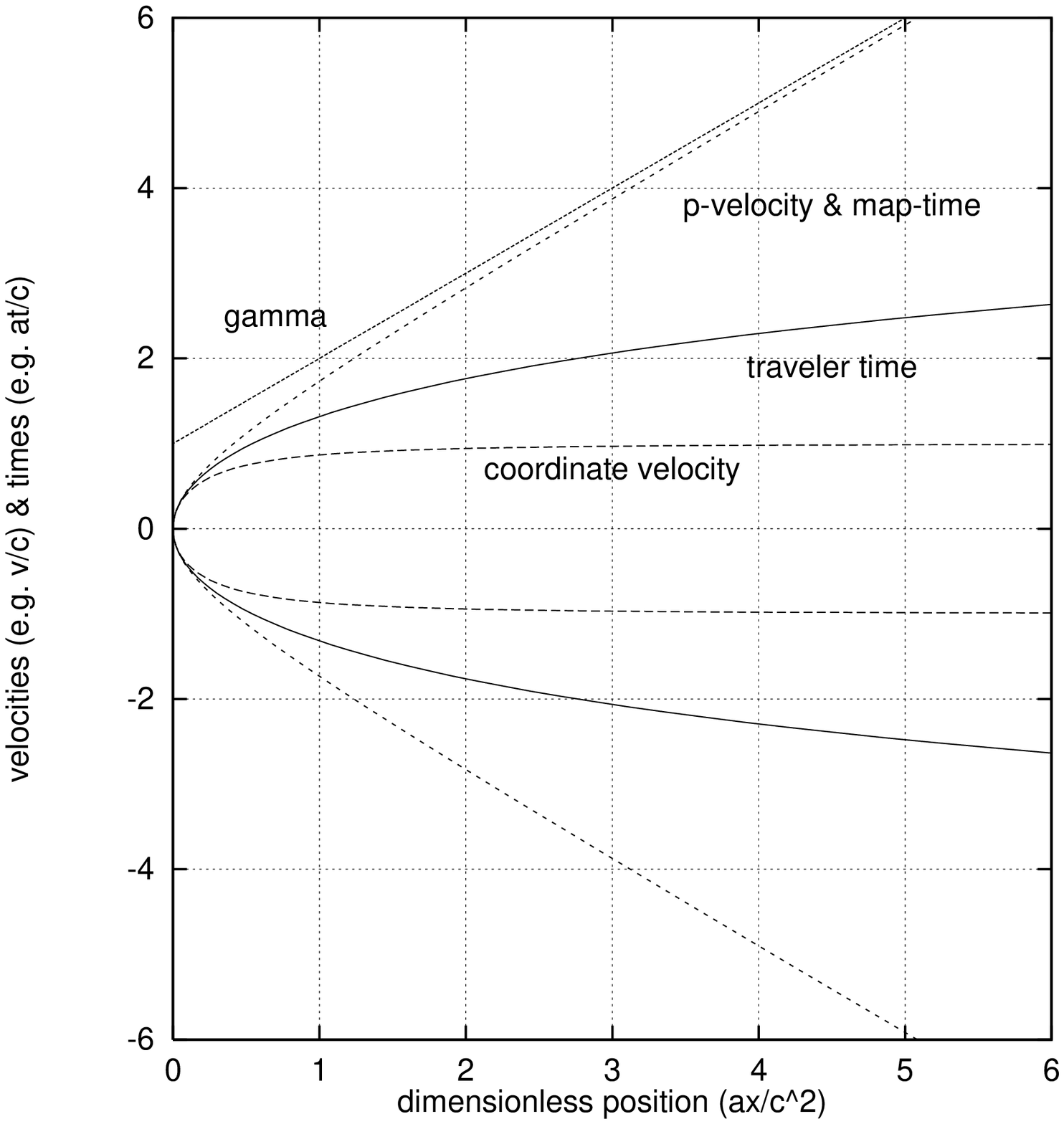}
\caption{The variables involved in (1+1)D constant acceleration}
\label{Fig1}
\end{figure}


\begin{references}
\bibitem{French}  e.g. A. P. French, {\em Special Relativity} (W. W. Norton,
NY, 1968), p.22.

\bibitem{Blatt}  e.g. F. J. Blatt, {\em Modern Physics} (McGraw-Hill, NY,
1992).

\bibitem{Adler}  C. G. Adler, Does mass really depend on velocity, dad?, 
{\em Amer. J. Physics} {\bf 55} (1987) 739-743.

\bibitem{Goldstein}  H. Goldstein, {\em Classical Mechanics, 7th printing}
(Addison-Wesley, Reading MA, 1965), p. 205.

\bibitem{SearsBrehme}  Sears and Brehme, {\em Introduction to the Theory of
Relativity} (Addison-Wesley, NY, 1968).

\bibitem{Shurcliff}  W. A. Shurcliff, {\em Special Relativity: The Central
Ideas} (19 Appleton St., Cambridge MA 02138, 1996).

\bibitem{Winnie}  J. A. Winnie, ``Special relativity without one-way
velocity assumptions, Part I and II'', {\em Philos. Sci.} {\bf 37} (1970)
81-99 and 223-228.

\bibitem{Ungar1}  A. A. Ungar, ``Formalism to deal with Reichenbach's
special theory of relativity'', {\em Found. Phys.} {\bf 21} (1991) 691-726.

\bibitem{Ungar2}  A. A. Ungar, ``Gyrogroup axioms for the abstract Thomas
precession and their use in relativistic physics and hyperbolic geometry'', 
{\em Found. Phys.} submitted (1996).

\bibitem{Noncoord}  P. Fraundorf, ``Non-coordinate time/velocity pairs in
special relativity'', {\em gr-qc/9607038} (xxx.lanl.gov archive, NM, 1996).

\bibitem{TaylorWheeler}  E. Taylor and J. A. Wheeler, {\em Spacetime
Physics, 1st edition} (W. H. Freeman, San Francisco, 1963).

\bibitem{LagouteDavoust}  C. Lagoute and E. Davoust, ``The interstellar
traveler'', {\em Am. J. Phys.} {\bf 63} (1995) 221.
\end{references}
\end{document}